\documentstyle[sprocl]{article}
\def\Journal#1#2#3#4{{#1} {\bf #2}, #3 (#4)}
\def\MPL{{\em Mod. Phys.Lett.} A}
\def\CMP{\em Comm. Math. Phys.}
\def\NPB{{\em Nucl. Phys.} B}
\def\PLB{{\em Phys. Lett.} B}
\def\PLA{{\em Phys. Lett.} A}

\def\l{\lambda}
\def\b{\beta}

\def\be{\begin{equation}}
\def\ee{\end{equation}}
\def\bea{\begin{eqnarray}}
\def\eea{\end{eqnarray}}

\begin{document}
\title{$R^2$ 2D Quantum Gravity and Dually Weighted Graphs
\footnote{ The talk is based on the recent paper of M.Staudacher,
T.Wynter and myself ~\cite{KSWIII} } }
\author{ Vladimir A. Kazakov }
\address{Laboratoire de Physique Th\'eorique  \\
de l'\'Ecole Normale Sup\'erieure
\footnote{Unit\'e Propre du
Centre National de la Recherche Scientifique,
associ\'ee \`a l'\'Ecole Normale Sup\'erieure et \`a
l'Universit\'e de Paris-Sud.}  \\
 24 rue Lhomond, 75231 Cedex 05, France}

\maketitle\abstracts{A recently introduced model of dually weighted
planar graphs is solved in terms of an .elliptic parametrization for some
particular collection of planar graphs describing the 2D $R^2$ quantum
gravity. Along with the cosmological constant $\lambda$ one has a
coupling $\beta$ in the model corresponding to the $R^2$-coupling
constant. It is shown that for any value of $\beta$ the large scale
behavior of the model corresponds to that of the standard pure 2D
quantum gravity. On small distances it describes the dynamics of
point-like curvature defects introduced into the flat 2D space. The
scaling function in the vicinity of almost flat metric is
obtained. The major steps of the exact solution are given.  }

\section{Introduction}

Matrix models counting the number of planar graphs of various
types have played an important role in various domains of physics and
mathematics, especially, in the quantitative approaches to 
two-dimensional quantum gravity and non-critical strings and string
field theories.

By means of a standard NxN hermitean 1-matrix with the partition function: 

 \begin{equation}
Z(t)=\int\,{\cal D}M\ e^{-{N\over 2} 
tr M^2 + N tr V(M)},
\label{eq:1mm}
\end{equation}
where the potential is defined by

 \begin{equation}
V(M)=
\sum_{k=1}^{\infty}{1\over k} t_k M^k ,
\label{eq:pot}
\end{equation}
we generate (by expanding $\log Z$ in powers of $t_k$) all the
abstract graphs weighted with the product of factors $t_k$ each
corresponding to a vertex with the coordination number $k$ and with
the overall factor $N^{2-2g}$ where $g$ is the genus of the graph. The
$t_k$'s allow us to control the frequencies of coordination numbers of
the vertices, but not the coordination numbers of the faces. One of
our main technical goals is to generalize the 1-matrix model
(~\ref{eq:1mm}) so that we can also weight the faces with independent
weights.
  
The model (\ref{eq:1mm}) has been proven to describe (with the
appropriate choices of $t_k$'s) the multicritical points of the pure
$2D$ gravity corresponding to the rational conformal matter (of the type
(2,2n-1) in the standard classification of the rational 2D conformal
theories) \cite{KAZ1,STAU}.

An interesting question to ask is whether the 1-matrix model could be
used for the description of the pure 2D gravity with higher derivative
terms, described by the following formal functional integral:

\begin{equation}
{\cal Z}(\lambda,\beta)= \int {\cal D}g_{ab}(z)~e^{-\int d^2z \sqrt{\det
g} {}~(\mu + \alpha R_g + {1\over \beta_0} R_g^2 + \ldots)}
\label{eq:cont}
\end{equation}

On the first sight, from the simple dimensional analysis, apart from
the cosmological term (controlled by $\mu$) and the (topological)
Einstein term proportional to the curvature $R_g$, it does not seem
meaningful to put further terms into the action of 2D gravity. The
simplest term one might want to consider is ${1 \over \beta_0}
R_{g}^{2}$. The bare coupling constant $\beta_0$ is however
dimensionful and thus should be proportional to the cutoff squared.
So it is small and in principle should be dropped, as well as any
further higher derivative terms.

On the other hand, if we make $\beta$ smaller and smaller the
characteristic metrics in the functional integral (~\ref{eq:cont})
should approach the flat one, since the $R^2$ suppresses the
fluctuations of the metric. So, an interesting question is whether one
can get some interesting non-perturbative behavior, where the $R^2$
coupling could play some essential role. In other words, the question
is whether there could be a non-perturbative phase of almost flat
metrics, beyond some hypothetic ``flattening'' phase transition. It is
clear that all the known approaches (like the Liouville theory
approach) which start from the continuum formulation (~\ref{eq:cont})
are not valid for this purpose since the formulation might need
serious modifications in this non-perturbative regime. Some
nonperturbative, lattice formulation is thus needed.

Unfortunately, the model (~\ref{eq:1mm}), in spite of the infinite
number of independent couplings, cannot be used to study a flattening
transition. It is impossible to tune the couplings in such a way that
we only generate flat, regular lattices corresponding to a flat
metric. The model (~\ref{eq:1mm}) provides no control over the
occurence of different types of faces, so even for only one non-zero
coupling, say, $t_k=\delta_{k,4}$, we get a sum over arbitrary
$\phi^4$ graphs, which describes the highly developed fluctuations of
pure 2d quantum gravity.

Hence we have to introduce a model where we can control the numbers of
faces with a given coordination number. The most general model of this
type, containing the second infinite set of dual couplings $t_k^*$,
which weight the faces (dual vertices) of graphs, is given by the
following partition function (here we fix the genus of the connected
graphs $g=0$):

 \begin{equation}
Z(t^*,t)=
\sum_{G} \prod_{v^*_q,v_q \in G} {t_q^*}^{\# v^*_q}\ {t_q}^{ \#
v_q} 
\label{eq:DWG}
\end{equation}

where $v_q^*,v_q$ are the vertices with $q$ neighbours on the
original
and dual graph, respectively, and $\# v_q^*,\# v_q$ are the numbers
of
such vertices in the given graph $G$. This expansion is
generated by the following matrix model:
\begin{equation}
Z(t^*,t)=\int\,{\cal D}M\ e^{-{N\over 2} tr~M^2\ +\ N tr~V_B(M A)},
\label{eq:DWGmatrix}
\end{equation}
with
\begin{equation}
V_B(M A)=
\sum_{k=1}^{\infty}{1\over k}~tr B^k\ (M A)^k .
\label{eq:potential}
\end{equation}
$A$ and $B$ are  fixed external matrices encoding the
coupling constants through
\begin{equation}
t_q^*={1\over q}\ {1\over N}\ tr\ B^q
{\rm \hskip 20pt and \hskip 20pt}
t_q={1\over q}\ {1\over N}\ tr\ A^q.
\label{eq:tqAB}
\end{equation}
The model generalizes, for $A \neq 1$, the standard one matrix model
(~\ref{eq:1mm}). We will call it the model of dually weighted graphs
(DWG).

The DWG model provides the possibility of flattening of the typical
graphs. If we take, say, $t_k=t_k^*=\delta_{k,4}$ we will single out
the regular square lattice in the partition function. The choice
$t_k=\delta_{k,3}$, $t_k^*=\delta_{k,6}$ brings us to the regular
hexagonal lattice, dual to the regular triangular lattice.  These
limits correspond to the $\beta=0$ limit in the continuum action of
(~\ref{eq:cont}). Our purpose here is to investigate the critical
behaviour of the DWG-model as we approach closer and closer this
limit.

\section{A Solvable Model of $R^2$ 2D Gravity} 

In the next section we will show that the DWG-model is in principle
solvable for any choice of coupling constants $\{t,t^*\}$. But for our
physical purposes we don't need so big space of independent
couplings. We have to find the simplest DWG-model with only two
independent couplings, one corresponding to the cosmological coupling
$\lambda$, another to the $R^2$ coupling $\beta$. 

We have found that the simplest model of this type corresponds to the
following choice of couplings:

\begin{equation}
t_2=\sqrt{\l}~t,~t_4=\l,~t_6=\l^{3 \over2}~{\beta^2 \over t},~...~
t_{2q}=\l^{q \over 2}~({\beta^2 \over t})^{(q-2)}.
\label{eq:weights}
\end{equation}

With these weights, it is easy to prove, using Euler's theorem, that
the partition sum ~\cite{KSWIII} becomes
\begin{equation}
{\cal Z}(t,\l,\b) = t^4~\sum_{G}~\l^A~\b^{2 (\# v_{2}-4)}, 
\label{eq:PART}
\end{equation}
where $A$ is the number of plaquets of the graph $G$ and $\#v_2$ the
number of positive curvature defects. Note that the latter are
balanced by a gas of negative curvature defects, whose individual
probabilities are given in (~\ref{eq:weights}).

After tuning the bare cosmological constant $\l$ (controlling the
number of plaquets) to some critical value $\lambda_c(\b)$, we expect
this model to describe pure gravity in a large interval of $\b$. On
the other hand, for $\l$ fixed and $\b=0$ we entirely suppress
curvature defects except for the four positive defects needed to close
the regular lattice into a sphere. It is thus clear that $\b$ is the
precise lattice analog of the bare curvature coupling $\b_0$ in
(~\ref{eq:cont}).  The phase $\b=0$ of ``almost flat'' lattices --
very different from pure gravity -- was discussed in detail in
~\cite{KSWII}.

\section{Sketch of Solution}

\subsection{The Itzykson-DiFrancesco formula for the DWG-model}

The basic fact which allows to solve the DWG-model is the
representation of the matrix integral (~\ref{eq:DWGmatrix}) in terms
of the character expansion with respect to the irreducible
representations R of the group $GL(N)$, given by Itzykson and
DiFrancesco ~\cite{IDiF}:

\begin{equation}
Z(t,t^*)=c\,\sum_{\{h^e,h^o\}}
{\prod_i(h^e_i-1)!!h^o_i!!\over
\prod_{i,j}(h^e_i-h^o_j)}~\chi_{\{h\}}(A)~\chi_{\{h\}}(B).
\label{eq:IDiF}
\end{equation}

The sum runs here over all representations $R\{h\}$ characterized by
the Young tableaux with $h_i-i+1$ boxes in the $i$-th row.  The
non-negative integers $h_i$ obey the inequalities:
\begin{equation}
h_{i+1}>h_i.
\label{eq:hconstr}
\end{equation}
 Only the representations with equal number of even(odd) weights
$h^e(h^o)$ enter the sum in (~\ref{eq:IDiF}).

The Weyl-Schur characters are defined in the standard way:

\begin{equation}
\chi_{\{h\}}(A)=\det_{_{\hskip -2pt (k,l)}}
                 \bigl(P_{h_k+1-l}(\theta)\bigr)
\label{eq:schuchar}
\end{equation}

as a determinant of Schur polynomials $P_n(\theta)$ defined through

\begin{equation}
e^{\Sigma_{i=1}^{\infty}z^i\theta_i}=\sum_{n=0}^{\infty}z^n P_n
(\theta)\quad {\rm with}\quad \theta_i={1\over i}~tr[A^i],
\label{eq:schupol}
\end{equation}

\subsection{The Saddle Point Equation for the Most Probable
Representation in the Planar Limit}

With the Itzykson-DiFrancesco formula we achieve what is necessary to
make the large N limit analytically treatable: the drastic reduction
of the number of degrees of freedom. Instead of $N^2$ integrations
over the hermitean matrix $M_{ij}$ we are left with only $N$
summations over the highest weight components $h_1<h_2<...<h_N$. If we
assume that the the characteristic $h's$ are of the order $N$ (which
will be confirmed by the solution) we can see that the summation
factor in (~\ref{eq:IDiF}) is of the order $\exp N$, and as usually
one can apply in this situation the saddle point approximation.

For our particular model (~\ref{eq:weights}-~\ref{eq:PART}) the saddle
point equation (SPE) is obtained by equating to zero the logarithmic
derivative of the weight of summation in (~\ref{eq:IDiF}) with respect
to $h^e$ or $h^o$ (we assume identical distributions for $h^e$ and
$h^o$).  The SPE reads:

\begin{equation}
2F(h)+P\int_0^a\ dh'\ {\rho(h') \over h-h'}= -\ln h.  
\label{eq:sdpt}
\end{equation}

where
\begin{equation}
F(h_k)=2{\partial \over \partial h^e_k}~\ln\ {\chi_{\{{h^e\over
2}\}}(a)
\over \Delta(h^e)}.
\label{eq:defF}
\end{equation}

and $\rho(h)$ is the density of the highest weight components in the
representation R. It is related to the resolvent  $H(h)$ of the
highest weight components:
\begin{equation}
H(h)=\sum_{k=1}^N {1\over h-h_k}=\int_0^a dh'\ {\rho(h') \over h-h'}.
\label{eq:res}
\end{equation}

Although the details of the derivation of the SPE contain some
subtleties, we can roughly say that the term $2F(h)$ comes from the
two characters in (~\ref{eq:IDiF}) normalized by the Van-der-Monde
determinant $\Delta(h)=\prod_{i>j}(h_i-h_j)$ (one takes $A=B$ for the
model dually equivalent to our model), the second term comes from log
derivative of $\Delta(h)$ and of the product
$\prod_{i,j}(h^e_i-h^o_j)$, and $\log h$ comes from the Sterling
asymptotics of $(h^e-1)!!$ (or $h^0!!$).

As has been discussed in detail in ~\cite{KSWI}, this equation
actually does not hold on the entire interval $[0,a]$, but only on an
interval $[b,a]$ with $0 \leq b \leq 1 \leq a$: Assuming the equation
to hold on $[0,a]$ would violate the implicit constraint $\rho(h) \leq
1$ following from the restriction (~\ref{eq:hconstr}).  The density is
in fact exactly saturated at its maximum value $\rho(h)=1$ on the
interval $[0,b]$.

A similar saddle point method for the representation expansion was
first successfully used in ~\cite{KD} for the calculation of the
partition function of the multicolour Young-Mills theory on the 2D
sphere (see also ~\cite{RU},~\cite{KW}).

\subsection{Calculation of Characters in the Large N Limit}

To solve the SPD (~\ref{eq:sdpt}) for $H(h)$ we have to find a method
to calculate effectively the characters, i.e. the function $F(h)$. In
this section we describe a method to do it, which is based on the
identities for Schur characters, based on the basic properties of
Schur polynomials. Leaving aside the details of the derivation (which
can be found in our basic paper ~\cite{KSWIII}), we summarize these
identities in one equation already in the large N limit. Introducing
the function:
\begin{equation}
G(h) = e^{H(h) +F(h)}
\label{eq:defG}
\end{equation}
we have:
\begin{equation}
h-1 = \sum_{q=1}^Q{t_{2 q}\over G^q} +
\sum_{q=1}^{\infty}
{2q\over N}{\partial\over \partial t_{2q}}\ln\Bigl(
\chi_{\{{h^e\over 2}\}}(a)\Bigr)
{}~G^q,
\label{eq:hGexp}
\end{equation}
where the coefficients of the positive powers of $G$ in
(~\ref{eq:hGexp}) are directly related to the correlators of the matrix
model dual to the model defined by the
eqs. (~\ref{eq:weights}-~\ref{eq:PART}), i.e. the model with potential
$V_{A_4}(\tilde M A) ={1\over 4}~(\tilde M A)^4$:
\begin{equation}
{2q\over N}{\partial\over \partial t_{2q}}\ln\Bigl(
\chi_{\{{h^e\over 2}\}}(a)\Bigr)=
\langle {1\over N} tr~(\tilde M A)^{2q} \rangle.
\label{eq:dualcorr}
\end{equation}

We have also assumed for the moment that only a finite number $Q$ of
couplings are non-zero (i.e.~$t_{2 q}=0$ for $q>Q$).  Furthermore, we
were able to show in our work ~\cite{KSWII} that (~\ref{eq:hGexp})
implies the functional equation
\begin{equation}
e^{H(h)}={(-1)^{(Q-1)}h\over t_Q}\prod_{q=1}^Q G_q(h),
\label{eq:HGprod}
\end{equation}
where the $G_q(h)$ are the first $Q$ branches of the multivalued
function $G(h)$ defined through (~\ref{eq:hGexp}) which map the point
$h=\infty$ to $G=0$. The saddlepoint equation (~\ref{eq:sdpt}),
together with (~\ref{eq:HGprod}), defines a well-posed Riemann-Hilbert
problem. It was solved exactly and in explicit detail in ~\cite{KSWII}
for the case $Q=2$, where the Riemann-Hilbert problem is succinctly
written in the form
\begin{equation}
\begin{array}{rcl}
 2 {\it Re} F(h)+H(h) &=&  -\ln (-{h\over t_4})   \\
 2F(h)+{\it Re} H(h)  &=& -\ln h,
\end{array}
\label{eq:twofph}
\end{equation}
where ${\it Re} H(h)$ denotes the real part of
$H(h)$ on the cut $[b,a]$ and ${\it Re} F(h)$ denotes the
real part of $F(h)$ on a cut $[-\infty,c]$ with
$c<b$. This case corresponds to an ensemble of squares being
able to meet in groups of four (i.e.~flat points) or two
(i.e.~positive curvature points). We termed the resulting surfaces
``almost flat''. It turned out that all the introduced functions
could be found explicitly in terms of elliptic functions.

The method can be easily generalized to any Q. However, the solution
cannot be in general expressed in terms of some known special functions.
In the next section we will show that the model of our interest can be
solved in terms of elliptic parametrization, more complicated than its
particular case, Q=2, of almost flat planar graphs.

Let us also note that the equation (~\ref{eq:HGprod}) can be explicitly
solved for $F(h)$ in terms of a contour integral for some wide class
of big Young tableaux. Hence we can find the character rather explicitly
knowing $H(h)$ and a set of Schur constants $(t_1,t_2,...,t_Q)$ (see
~\cite{KSWIII}).
 
\subsection{Solution of the Lattice Model of $2D$ $R^2$ Gravity}

Let us notice that for the model defined by
(~\ref{eq:weights},~\ref{eq:PART}) which we chose as a lattice
realization of the $2D$ $R^2$ gravity, the equation (~\ref{eq:hGexp})
contains an infinite number of both negative and positive powers of
the $G$-expansion.

Labeling the first two weights as in ~\cite{KSWII}: $t_2$, $t_4$, we
now include all the even $t_{2q}$ with $q\geq 3$, assigning them the
following weights: $t_{2q}=t_4\epsilon^{q-2}$.  Equation
(~\ref{eq:hGexp}) can then be written compactly as:
\begin{equation}
h-1 ={t_2\over G}+{t_4\over G~(G-\epsilon)}+{\rm\ positive\ powers\
of\ } G.
\label{eq:hGfull}
\end{equation}

Dropping the positive powers of $G$ (for $G$ small enough)
 and inverting this equation, we
see that there are two sheets connected together by a square root cut
running between two finite cut points, $d$ and $c$. On
the physical sheet a further cut (running from $b$ to $a$),
corresponding to $e^{H(h)}$, connects to further sheets. 

The equations (~\ref{eq:twofph})  now read:
\begin{equation}
\begin{array}{rcl}
 2{\it Re} F(h)+H(h) &=& -\ln ({h \over \epsilon t_2-t_4})     \\
 2F(h)+{\it Re} H(h) &=& -\ln h,
\end{array}
\label{eq:twofphmod}
\end{equation}
the first coming from the large $N$ limit of the character
(~\ref{eq:HGprod}) and the second, in view of (~\ref{eq:res}), being
the saddlepoint equation (~\ref{eq:sdpt}). These two equations tell us
about the behaviour of the function $2F(h)+H(h)$ on the cuts of $F(h)$
and $H(h)$, respectively. We have introduced the notation ${\it Re}
F(h)$ to denote the real part on the cut of $F(h)$, and similarly for
${\it Re} H(h)$. The principal part integral in (~\ref{eq:sdpt}) is
thus denoted in (~\ref{eq:twofphmod}) by ${\it Re} H(h)$.

Our object is to now reconstruct the analytic function
$2F(h)+H(h)=2\ln G(h)-H(h)$ from its behaviour on the cuts. To do this
we first need to understand the complete structure of cuts. We already
know the structure of cuts of $H(h)$; it has a logarithmic cut running
from $h=0$ to $h=b$, corresponding to the portion of the density which
is saturated at its maximal value of one, and a cut from $b$ to $a$
corresponding to the ``excited'' part of the density, where the
density is less than one. It thus remains for us to understand the cut
structure of $\ln G(h)$.

The function $G(h)$ has two cuts on the physical sheet. The first cut,
running from $b$ to $a$, corresponds to the cut of $e^{H(h)}$, the
second cut, running from cut point $c$ to cut point $d$, corresponds
to the cut of $e^{F(h)}$. To see whether $\ln G(h)$ has any
logarithmic cuts, we first notice from (~\ref{eq:hGfull}) that $G(h)$
is non zero everywhere in the complex $h$ plane except possibly at
infinity. Thus for $\ln G(h)$ the only finite logarithmic cut points
are at $h=b$, defined to be the end of the flat part of the density
(this corresponds to the end of the cut of $e^{H(h)}$), and possibly
the cut point $c$, defined to be at the end of the cut of
$e^{F(h)}$. The only remaining question is whether this logarithmic
cut starting at $h=b$ goes off to infinity or terminates at $c$. For
large $h$ we see from (~\ref{eq:hGfull}) that $\ln
G(h)=\ln\epsilon+{\cal O}({1\over h})$, i.e. there is no logarithmic
cut at infinity. We conclude that the cut structure of the function
$\ln G(h)$ consists of the cuts corresponding to $e^F(h)$ and
$e^{H(h)}$ connected together by a logarithmic cut, whose cut points
are $b$ and $c$.

We thus understand the behaviour of $2 F(h) + H(h)$ on all of its
cuts. Standard methods then allow us to generate the full analytic
function $2 F(h) + H(h)$. The four cut points, $a$ and $b$ defining
the cut of $ H(h)$, and $c$ and $d$ defining the cut of $F(h)$ are
fixed later by boundary conditions.

Without going into the details of the solution let us present at least
one final result: the density of weights $\rho(h)$ of the most
probable representation $R\{h\}$. It can be obtained from $2F(h)+H(h)$
by using the saddle point equation $2F(h)+{\it Re} H(h)=-\ln h$ and
the fact that the resolvent (~\ref{eq:res}) for the Young tableau can
be written as $H(h)={\it Re} H(h) \mp i\pi\rho(h)$. To obtain the
result one has to use all the analytical information listed above.
After some tedious calculations we obtain:
\begin{equation}
\rho(h)=
{u \over K} - {i \over \pi}~\ln \Bigg[
{\theta_4\big({\pi \over 2 K} ( u - i v),q \big)
\over
 \theta_4\big({\pi \over 2 K} ( u + i v),q \big)} \Bigg], 
\label{eq:density}
\end{equation}
where $v$ and $u$ are defined by
\begin{equation}
v=sn^{-1}\bigl(\sqrt{{a-c \over a-d}},k'\bigr), \ \ \ \ \
u=sn^{-1}\bigl(\sqrt{{(a-h)(b-d) \over (a-b)(h-d)}},k\bigr),
\label{eq:defvu}
\end{equation}
and
\begin{equation}
k^2={(a-b)(c-d) \over (a-c)(b-d)}.
\label{eq:defk}
\end{equation}
$K$ and $K'$ are the complete elliptic integrals of the first kind
with respective moduli $k$ and $k'=\sqrt{1-k^2}$. $E$ is the complete
elliptic integral of the second kind with modulus $k$, $E(v,k')$ is
the incomplete elliptic integral of the second kind with argument $v$
and modulus $k'$ and $sn$, $cn$ and $dn$ are the Jacobi Elliptic
functions. The nome $q$ is defined by
\begin{equation}
q=e^{- \pi {K' \over K}}.
\label{eq:nom}
\end{equation}

To fix the constants $a$, $b$, $c$ and $d$, we expand $i
\pi\rho(h)=2F(h)+H(h)+\ln h$ for large $h$ and compare the resulting
power series expansion to that obtained from inverting
(~\ref{eq:hGexp}):
\begin{equation}
2F(h)+H(h)=2\ln\epsilon+\bigl({2t_4\over\epsilon^2}-1\bigr){1\over h}
\quad+\quad{\cal O}\bigl({1\over h^2}).
\label{eq:bcexpn}
\end{equation}
The terms of ${\cal O}\bigl({1\over h^2})$ depend on the as yet
unknown positive powers of $G$ in (~\ref{eq:hGfull}).

Afinal boundary condition is fixed by the normalization of the
density.

Using this information we can calculate the physical quantities of
interest, for example, the derivative of the free energy with respect
to the cosmological constant.  Denoting the free energy by ${\cal
Z}(t,\lambda,\beta)$, $Z=e^{-N^2{\cal Z}}$, we have
\begin{equation}
{\partial\over\partial\lambda}{\cal Z}(t,\lambda,\beta)=
      {1\over 4\lambda}(\langle{1\over N} tr M^2\rangle -1).
={1\over 2}+\langle h \rangle
\label{eq:dlfem}
\end{equation}

where $\langle h \rangle = \int dh~\rho(h)~h$. The result of the last
integration, as well as the conditions for $a,b,c,d$ are too bulky to
be presented here (see ~\cite{KSWIII} for the details). In the next
section we will present already the physical consequences of it,
namely, the universal expression for the free energy near the
flattening limit. 

\section{Physical Results and Conclusions}

One needs to do quite tedious calculations to extract the physical
results from our solution summerized in the equations
(~\ref{eq:density}-~\ref{eq:dlfem}). The details can be found in
\cite{KSWIII} and here we present only the essential results:

1. We found the equation for the critical curve of pure gravity in the
parameter space $(\lambda,\beta)$. It did not show any ``flattening''
phase transition in the whole interval $0<\beta<\infty$. Pure gravity
appears to be the only possible critical behaviour for the large scale
fluctuations of the metric (the only infrared stable critical point)
within our model. This should be true for a big class of similar
lattice models of pure $R^2$ gravity in 2 dimensions. For example, if
we would allow only the deficits of angle being $0,\pm \pi$ in the
vertices of our quadrangulation, this physical conclusion should not
change. On the other hand, the multicritical behaviours can occur for
a more complex parameter space (more $t^*,t$ couplings to vary) and we
could expect some new phase structures. Sums over graphs with the
sign-changing Boltzmann weights might well be necessary for this.

2. Once we realize that nothing interesting happens in our model for
finite values of $\beta$ we should concentrate our attention on the
double limit $\b \rightarrow 0, \lambda \rightarrow
\lambda_c(\beta)=1-{\pi \over \sqrt{2}}\b + O(\beta^2)$ with the
double limit parameter $x=1+{\sqrt{2} \over \pi \b}
(\lambda_c-\lambda)$ fixed. Note that this is a very natural,
dimensionless scaling parameter; $\lambda_c-\lambda$ is the continuum
cosmological constant with dimension of inverse area and $\beta$
controls the number of curvature defects per unit area (and thus also
has dimension of inverse area).

We arrive at the free energy in the vicinity of the
(locally) flat metric:
\begin{equation}
{\cal Z}(t,\lambda,\beta)=
{4 t^4\over 15\beta^2}\bigl[x^6-{5\over 2}x^4+{15\over 8}x^2-{5\over
16}
-x(x^2-1)^{5/2}\bigr].
\label{eq:feflat}
\end{equation}

This scaling function exhibits two different behaviours in the two
opposite limits:

1. $x_c \rightarrow 1$ - the pure 2D gravity regime for any finite
positive $\beta$ and $\lambda \rightarrow
\lambda_c(\beta)$:
\begin{equation}
{\cal Z}(t,\lambda,\beta) \sim (\lambda_c-\lambda)^{5/2}
\label{eq:pureg}
\end{equation}
This corresponds to the standard value of the string susceptibility
exponent $\gamma_{str}=-1/2$.

This results mean that even for very big but finite $R^2$-coupling
$\beta^{-1}$ we will always find on big enough distances the metric
fluctuations obeying the pure gravity scaling. In some sense one can
say, that the $R^2$ 2D gravity does not exist as a special phase of
the 2D gravity. The $R^2$ term can be always dropped from the action
in the long wave limit. Our results demonstrate this
nonperturbatively, starting from a solvable lattice model.

2. In the opposite limit $x\rightarrow \infty$, i.e., $\beta
\rightarrow 0$ with $\lambda$ fixed, we obtain:

\begin{equation}
{\cal Z}(t,\lambda,\beta) = {\pi^2\over48} {\beta^4 \over (1-\lambda)^2}
\label{eq:dilga}
\end{equation}

This limit corresponds to a sum over metrics flat everywhere except of
a finite number of curvature defects. One can easily see from the
purely combinatorial calculations that (~\ref{eq:dilga}) corresponds
to the sum over all flat metrics with only four defects with the
deficit of angle $\pi$. Expanding (~\ref{eq:feflat}) as a power series
in $\b$ (the first term of which is given by (~\ref{eq:pureg}) we see
that the second term in the series expansion corresponds to the the
sum over all flat metrics with five $\pi$ defects plus one $-\pi$
defect, the third to six $\pi$ defects plus two $-\pi$ defects, etc.

This limit does not appear to describe a new phase of smooth
metrics. It corresponds instead to the statistics of a dilute gas of
curvature defects introduced in an otherwise flat metric. It has
nothing to do with the 2D gravity but it is an interesting statistical
mechanical model in itself. More than that, one can prove that the
dominating graphs are not lattice artifacts: in this limit the
manifold ``forgets'' that it is built out of squares. It consists of
big flat patches with sparsely distributed point-like curvature
defects introduced ono it in such a way that they close the surface
into a manifold of spherical topology. If one considers this object as
a sphere punctured at the points where curvature defects occur, one
can show by the direct calculations\footnote{we thank Paul Zinn-Justin
for clarifying this question for us}, for the simplest configurations,
that the corresponding moduli parameters do not fall on the boundaries
of the moduli space but are typically in some general position inside
the fundamental domain.

It would be nice to solve this model (and may be more general cases
including the matter fields) by a continuum approach. In the Liouville
theory picture the curvature defects correspond to coulomb charges
floating in a two-dimensional parametrization space. This could be an
alternative approach to 2D quantum gravity in general.

Let us conclude by noting that our analytical approach to the models
of dually weighted graphs can be generalized to a big class of other
matrix models which could be of physical, as well as of the
mathematical interest. For example, one can formulate a generalized
two-matrix model with partition function:

\begin{equation}
Z=\int d^{N^2}X d^{N^2}Y \exp N tr( F(XY) + U(X) +V(Y) )
\label{eq:2mm} 
\end{equation}
where $X,Y$ are the $N$x$N$ hermitean matrices and $F,U,V$ are some
arbitrary functions. The character expansion allows us again to reduce
the number of integration (or summation) variables to $\sim N$ thus
making it possible to consider it by means of the saddle point
methods. Another, more general solvable k-matrix model of this type
is:

\begin{equation}
Z=\int \big(\prod_{j=1}^k d^{N^2}X_j\big)  \exp N tr\big( F(\prod_{j=1}^k
X_j) + \sum_{j=1}^k U_j(X_j) \big)
\label{eq:kmm}
\end{equation}

The $\exp N tr F(\prod_{j=1}^k X_j)$ factor can be also expended in
characters and one can then integrate over the angular degrees of
freedom of the matrices by the use of only the simple
SU(N)-orthogonality relations. This approach is at the heart of the
Itzykson-DiFrancesko formula which we extensively used for our present
model of lattice 2D $R^2$ gravity.

A big question left is how to generalize our approach to include of
the matter fields on our lattice manifold. The corresponding matrix
models containing the $R^2$ type couplings can be easily formulated
but cannot be solved by the character expansion methods presented
here. This is not surprising, since models of interacting spins on the
2D regular lattice, even the integrable ones, are very
complicated. Nevertheless, we think that our methods could eventually
be powerful enough to advance in this direction and provide a missing
link between two branches of mathematical physics: integrable
statistical mechanical models on regular 2D lattices, on the one hand,
and on random dynamical lattices, on the other hand. In more physical
language, it could provide a link between the integrable models of
interacting fields with and without the 2D quantum gravity
fluctuations.


\begin{thebibliography}{99}
\bibitem{KSWIII} V.A.~Kazakov, M.~Staudacher and T.~Wynter, \'Ecole
Normale preprint LPTENS-95/56, hep-th/96011069, accepted for
publication in ~Nucl.~Phys. B
\bibitem{KAZ1} V.A.~Kazakov, \Journal{\MPL}{4}{1691}{1989}.
\bibitem{STAU} M.~Staudacher, \Journal{\NPB}{336}{349}{1990}.
\bibitem{KSWI} V.A.~Kazakov, M.~Staudacher and T.~Wynter, \'Ecole
Normale preprint LPTENS-95/9, CERN-TH/95-352, hep-th/9502132, 
to be published in \Journal{\CMP}{}{}{}.
\bibitem{KSWII} V.A.~Kazakov, M.~Staudacher and T.~Wynter, Ecole
Normale preprint LPTENS-95/24, hep-th/9506174, 
to be published in \Journal{\CMP}{}{}{}.
\bibitem{KAW} H.~Kawai and R.~Nakayama,
\Journal{\PLB}{306}{224}{1993}.
\bibitem{KD} M.~Douglas and V.~Kazakov, \Journal{\PLB}{319}{219}{1993}. 
\bibitem{RU} B.~Rusakov, \Journal{\PLB}{303}{95}{1993}.
\bibitem{KW} V.A.~Kazakov and T.~Wynter, \Journal{\NPB}{440}{407}{1995}.
\bibitem{BIPZ} E.~Brezin, C.~Itzykson, G.~Parisi and J.B.~Zuber,
\Journal{\CMP}{59}{35}{1978}.
\bibitem{K} V.~Kazakov, \Journal{\PLA}{119}{140}{1986}.
\bibitem{IDiF} P.~Di~Francesco and C. Itzykson, Ann. Inst. Henri.
Poincar\'e Vol. 59, no. 2 (1993) 117.




\end{thebibliography}
\end{document}